\documentclass[Times,twocolumn,tighten,deluxetable]{aastex63}

\usepackage{graphicx}
\usepackage{lineno}
\DeclareGraphicsExtensions{.pdf,.jpeg,.png}

\submitjournal{ApJ}

\shorttitle{Outburst Profiles of 4U 1630-472}
\shortauthors{K. Chatterjee et al.}

\begin{document}

\title{Anomalous nature of outbursts of the black hole candidate 4U 1630-472}

\correspondingauthor{Dipak Debnath}
\email{dipakcsp@gmail.com}

\author[0000-0002-6252-3750]{Kaushik Chatterjee}
\affiliation{Indian Center for Space Physics, 43 Chalantika, Garia St. Road, Kolkata 700084, India}

\author[0000-0003-1856-5504]{Dipak Debnath}
\affiliation{Indian Center for Space Physics, 43 Chalantika, Garia St. Road, Kolkata 700084, India}

\author[0000-0002-7658-0350]{Riya Bhowmick}
\affiliation{Indian Center for Space Physics, 43 Chalantika, Garia St. Road, Kolkata 700084, India}

\author[0000-0002-6640-0301]{Sujoy Kumar Nath}
\affiliation{Indian Center for Space Physics, 43 Chalantika, Garia St. Road, Kolkata 700084, India}

\author[0000-0001-6770-8351]{Debjit Chatterjee}
\affiliation{Indian Institute of Astrophysics, Koramangala, Bengaluru, Karnataka, 560034, India}
\affiliation{Indian Center for Space Physics, 43 Chalantika, Garia St. Road, Kolkata 700084, India}





\begin{abstract}

The Galactic black hole candidate (BHC) 4U~1630-472 has gone through several outbursts (13 to be particular) in the 
last two and a half decades starting from the RXTE era till date. Like the outbursts of other transient BHCs, 
the outbursts of this source show variations in duration, peak numbers, highest peak flux, etc. However, unlike any
other soft X-ray transients, this source showed outbursts of two types, such as normal and super. The normal outbursts 
of duration $\sim 100-200$~days are observed quasi periodically at an average recurrence/quiescence period of $\sim 500$~days. 
The super outbursts of duration $\sim 1.5-2.5$~years contain one or more normal outbursts other than one mega outburst. 
We make an effort to separate flux contribution of the normal and the mega outbursts from the super outbursts, and 
tried to understand the nature of evolution of both types (normal and mega) of outbursts, based on the quiescent 
period prior to the outbursts. Archival data of RXTE/ASM from January 1996 to June 2011, and MAXI/GSC from August 2009 
to July 2020 are used for our study. A possible linear relation between the quiescent and outburstsing periods for both 
types of outbursts are observed. This makes the BHC a special source, and it may contain two companion binaries. 
Two companions might be responsible for two types of outbursts. 

\end{abstract}


\section{Introduction}

Compact objects like white dwarfs (WDs), neutron stars (NSs) and black holes (BHs) are the end products of stars. Astrophysical black holes form from 
more massive stars than those which end up their lives as WDs or NSs. Based on the mass, BHs are divided into three classes: (a) stellar mass black 
holes (SBHs), (b) intermediate mass black holes (IMBHs) and (c) super massive black holes (SMBHs). Depending on the spin, BHs are divided into two 
categories: (a) non-rotating Schwarzschild black holes and (b) rotating Kerr black holes. The most prominent way to detect a BH is by detecting the 
radiation that is coming from the accretion disk of the BH during the course of an outburst. Outbursts are thus of very much importance in case of 
detection of a BH. We are interested about the outburst of the stellar mass black holes. Depending on the outburst nature, they are mainly of two types: 
transient and persistent. Transient black holes are those which mostly stay in the quiescent phase and occasionally become active and show outbursts 
when there is rise-in viscosity of the infalling matter from the companion. Mostly we are interested in these transient SBHs and their outbursts. 
The radiation from the close region of the disk of these outbursts are in X-rays. Generally, for most of the times, two different outbursts of the same
BH are likely to be different. The nature of the light curves of outbursts of BH transients divide themselves into various different kinds of morphologies,
of which, according to Chen et al. (1997), the most prominent one is the fast rise exponential decay (FRED, Kocevski et al. 2003) type. Debnath et al. 
(2010), classified BH sources based on their outburst profiles into two classes: fast-rise slow-decay (FRSD) and slow-rise slow-decay (SRSD). During an 
outburst of a transient black hole, we observe transition between various spectral states. There are four defined spectral states of a black hole 
candidate (BHC): hard state (HS), hard inter-mediate state (HIMS), soft inter-mediate state (SIMS) and soft state (SS) (Remillard \& McClintock 2006). 
Generally, during an outburst, a transient BHC makes spectral state transition as HS $\rightarrow$ HIMS $\rightarrow$ SIMS $\rightarrow$ SS $\rightarrow$ 
SIMS $\rightarrow$ HIMS $\rightarrow$ HS. This type of outbursts are called type-I outburst. In case of type-II outburst, we do not see transition towards 
the softer states and these outbursts are also known as `failed' outbursts (Debnath et al. 2017 and the references therein). Low frequency quasi periodic 
oscillations (LFQPOs) are also the common feature in most of the BH outbursts, which are mainly observed in the HS and HIMS. LFQPOs are also observed 
in SIMS, but sporadically on and off (Casella et al. 2005; van der Klis 2005). Generally, it has been observed that during the rising phase of the outburst, 
the QPO frequency rises monotonically while it decreases in the declining phase (Nandi et al. 2012; Debnath et al. 2013). LFQPOs are generally absent in SS.

The soft X-ray transient (SXT) 4U 1630-472 (Kuulkers et al. 1997) was discovered by the \textit{Uhuru} satellite (Jones et al. 1976). Based on its spectral
and temporal behaviour, 4U 1630-472 is thought to be the repository of a black hole (Kuulkers et al. 1997).  This is a very active SXT with occurrence of 
outburst at roughly every 400-550 days (Jones et al. 1976; Priedhorsky 1986). Its first outburst was recorded back in 1969 by \textit{VELA 5B} (Priedhorsky 
1986) and since then it has gone through outbursts more than 20 times. With this recurrence time, 4U 1630-472 is known to be the shortest recurrent SXT
(Kuulkers et al. 1997). According to Seifina et al. (2014), it has a mass $\sim 10~M_{\odot}$ and is at a distance of $10-11~kpc$. There are presence of 
three pulsars, such as IGR~J16358-4726, IGR~J16393-4643, and IGR~J16320-4751, in very close angular separation of 4U~1630-472. These sources are visible 
in the ASM and GSC field of views (FOVs) of 4U~1630-472. There could be contamination of these sources due to this close separation during the outbursts 
of 4U 1630-472. Due to the crowded region where it is situated, no companion has been clearly detected yet (Kuulkers et al. 1997). However, from its 
IR properties, Capitanio et al. (2015) suggested that it contains an early-type secondary star with an orbital period ($P_{orb}$) of a few days. According 
to Parmar et al. (1997), the outbursts of this SXT is not strictly periodic, but vary by an $8~\%$ of rms. 

There has always been this debate about the mechanism that triggers an outburst. Primarily, it is due to the fact that when there is enhancement in viscosity,
matter comes closer to the compact object, triggering an outburst. Chakrabarti et al. (2019, hereafter CDN19) tried to explain the triggering mechanism in
black holes by analysing the outburst profiles of the BHC H 1743-322. Bhowmick et al. (2021) studied relation between quiescent and outburst phases of the 
well known Galactic transient BHC GX~339-4 in the post RXTE era. These works motivated us to study outburst profiles of 4U 1630-472 as it is one of the most 
active transient BHCs. It shows outbursts in a frequent interval of times. In the post RXTE era (1996-2020), it showed 13 outbursts, out of them 3 are of long 
duration and rest are of short duration. 

This paper is organized as the following. In \S2, we present the analysis technique. In \S3, we will present our results. Finally in \S4, we will discuss about 
the results and concluding remarks will be made.

\section{Analysis}

4U 1630-472 is a highly active recurring transient BHC, which shows outbursts frequently. Light curve profiles from the RXTE to post-RXTE eta (1996-2020) are 
used for the study in the paper. We used RXTE/ASM data (1996-2011) in the $1.5-12~keV$ and MAXI/GSC data (2009-2020) in $2-10~keV$ energy bands for our study. 
2009--10 outburst is commonly observed by both the instruments (ASM and GSC). So, it allowed us to normalize integrated fluxes and other measured 
quantities from the data of the two instruments. We downloaded archival daily average light curves which are in $counts/s$ and $photons~cm^{-2}~s^{-1}$ units in 
case of ASM and GSC data respectively. With proper conversion factor, we have converted all the light curves into $mCrab$ unit. For ASM, $1~mCrab$ 
equals $0.075~counts/s$ in the $1.5-12~keV$ energy band while, in case of GSC, $1~mCrab$ is equal to $0.00282~photons~cm^{-2}~s^{-1}$.

It is mentioned earlier that the most prominent morphology amongst the outbursting light curve profiles is FRED. Here, we also can notice that all the 
light curve profiles contain multiple peaks, which are FRED in nature. For our current work, a profile fitting of the light curves are necessary. 
It helped us to measure properties (peak flux, peak time, duration integral flux, etc.) of each outburst. To do this, we make use of the FRED model 
(will be discussed in later Section). Using this model, we used a code written in C++ language, which takes care of the FRED morphology of light curve 
profiles. We fit all the outburst profiles (shown in Figure 1) using one or combination of multiple FRED profiles. For fitting with this code, we make 
use of CERN's data analysis software, ROOT (version 6.13/01). We also calculated integral flux of the outbursts using simple integration method. 
It helped us to verify FRED fitted results.

We have also estimated the periodicity of the duration and gap between normal outbursts using NASA's 
\href{https://exoplanetarchive.ipac.caltech.edu/cgi-bin/Pgram/nph-pgram}{Lomb Scargle Periodogram} website (Peter et al. 2008; Kovacs et al. 2002; Horne
\& Baliunas 1986; Scargle 1982).

To check for the contamination from nearby sources, we have also downloaded daily average light curves of the IGR pulsar sources from Swift/BAT, RXTE/ASM 
and MAXI/GSC archives and fitted each available curves with FRED profile by converting them into mCrab unit.

\section{Results}

\begin{figure*}
\vskip 1.2cm
  \centering
    \includegraphics[angle=0,width=11cm,keepaspectratio=true]{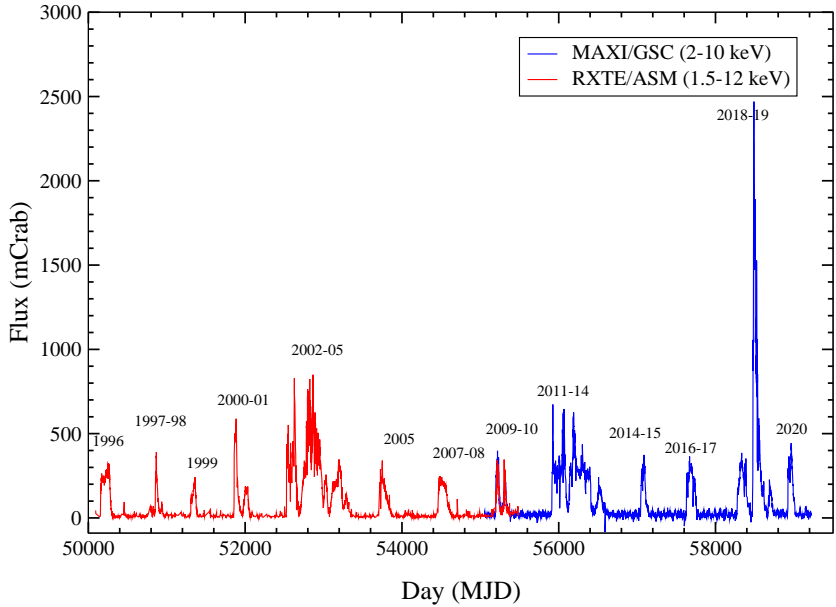}
    \caption{Daily averaged and mCrab converted light curve profiles of various outbursts of the BHC 4U 1630-472. The online red curves are $1.5-12~keV$
             RXTE/ASM data and online blue represent the $2-12~keV$ MAXI/GSC data. All the light curve data are converted into $mCrab$ unit using proper
             conversion factors.}
\end{figure*}

Comparative study of light curves and duration of the quiescent period prior to the outbursts, help us to understand the nature of the outbursts along 
with the other underlying properties. It helps us to understand the triggering mechanism of the outbursts along with the relation of the outbursts 
with their preceding quiescent periods. To do this, we have fitted a total of 13 outbursts of 4U 1630-472 using the FRED model in ROOT. Out of those, 
8 outbursts are from RXTE era and we have made use of ASM data. The rest 6 are from post-RXTE era and we have taken MAXI/GSC data for those. The 2009--10 
outburst is common for both the instruments. It helped us to normalize integral flux of other outbursts. If Fig. 1, we have shown daily average fluxes in 
$mCrab$ unit. 

In the quiescent period between two outbursts, photon counts/fluxes in the light curves stay at very low range. We define start and end of an outburst 
by taking $12~mCrab$ flux as the threshold. This threshold was also used by CDN19 and Bhowmick et al. (2021) in their studies for the BHCs H~1743-322 and 
GX~339-4 respectively. In Fig. 2, to signify this threshold, two vertical lines have been drawn other than one horizontal line at $12~mCrab$ flux for all
four shown figures. Detailed analysis results are mentioned in the following sub-Sections.

\subsection{Outburst Profiles}

In Figure 1, we show the daily average light curve profiles (in mCrab) for the BHC 4U 1630-472 in the RXTE and MAXI era. RXTE/ASM light curve in the $1.5-12~keV$ 
energy band from $1996$ to $2009-10$ is shown with the red curve, while $2-10~keV$ MAXI/GSC light curve from $2009-10$ to $2020$ is shown with the blue curve. 
We see $8$ outbursts with the ASM data and $6$ outbursts with the GSC data. The $2009-10$ outburst was a commonly observed outburst by the two instruments. 
We used integrated flux and other parameters obtained from this outburst for the normalization to the other outbursts observed lonely by the two instruments. 
In these total $13$ outbursts a large variation in the peak flux, duration, etc are observed. If we compare these outbursts with outbursts of other SXTs, these 
outbursts of 4U 1630-472 are very much peculiar and unfamiliar in nature. In general, an SXT shows outbursts every $2-3$ years for the duration of a few months 
of active periods. But 4U~1630-472 showed two types of outbursts with differences in duration, integrated flux, prior quiescent period, etc.

4U 1630-472 is a very interesting transient BHC as it showed outbursts in a short gap of the recurrence period. From the RXTE era (1996-2020), 4U 1630-472 
has gone through 13 outbursts. Most of its outbursts lasted for almost $100-200$ days while some has lasted more than $500$ days. 
We can classify 4U 1630-472 outbursts into two types: \textit{normal} and \textit{super}. The normal outbursts are recurring and observed in a gap of 
$\sim 400-550$ days. These outburst duration roughly followed a quasi-periodic variations, if we looked into the structure of the \textit{super outbursts}. 
It seems that super outbursts contain one or two \textit{normal-outbursts} other than one \textit{mega-outburst}.

The super outburst that started in early September 2002 (MJD $52521.4 \pm 0.4$), and ended in December 2004 (MJD $53368.7 \pm 22.0$), i.e., lasted for 
$\sim 2.35$ years ($\sim 850$ days). This was in the RXTE era. In the MAXI era too, there are two super outbursts. One of which started in late November 
2011 (MJD $55889.7 \pm 6.2$) and has continued till the very end of December 2013 (MJD $56651.2 \pm 15.0$), lasting for $\sim 2.1$ years ($\sim 760$ days). 
The other occurred in 2018--19, which started in April 2018 (MJD $58233.9 \pm 1.2$) and came below threshold in November 2019 (MJD $58789.2 \pm 6.9$), lasting
for almost 1.5 years ($\sim$ 550 days). Since, there is some error in the value at which date the $\sim 12~mCrab$ flux has been reached, especially in the 
declining phase, we have taken the first super outburst as $2002-05$ instead of $2002-04$ and the second one as $2011-14$ instead of $2011-13$. There is also 
various reports of some super outbursts before the RXTE era, for example the 1988 outburst, that lasted for $\sim 2.4$ years (Kuulkers et al. 1997). Before 
that there were also report about the super outburst in 1977. 

Although the average recurrence period for normal outbursts is close to $500$ days, the outbursts are not always exactly periodic. For example, the 1997--98 
outburst occurred after a recurrence period of $\sim$ 540 days, while the next outburst in 1999 occurred after a period of $\sim$ 400 days. The duration of 
outbursts also differ from one another. In case of the super outburst also, this duration is changing. In fact, if we notice carefully, we find that the 
super outbursts are becoming smaller in duration. While the super outburst in 1988-91 lasted for almost 870 days (Kulkeers et al. 1997), we find that the 
2002--05, 2011--14 and 2018--19 have duration of $\sim$ 850, 760 and 550 days respectively.

\begin{figure}
\centering
\vbox{
\includegraphics[width=8.0truecm, height=4cm]{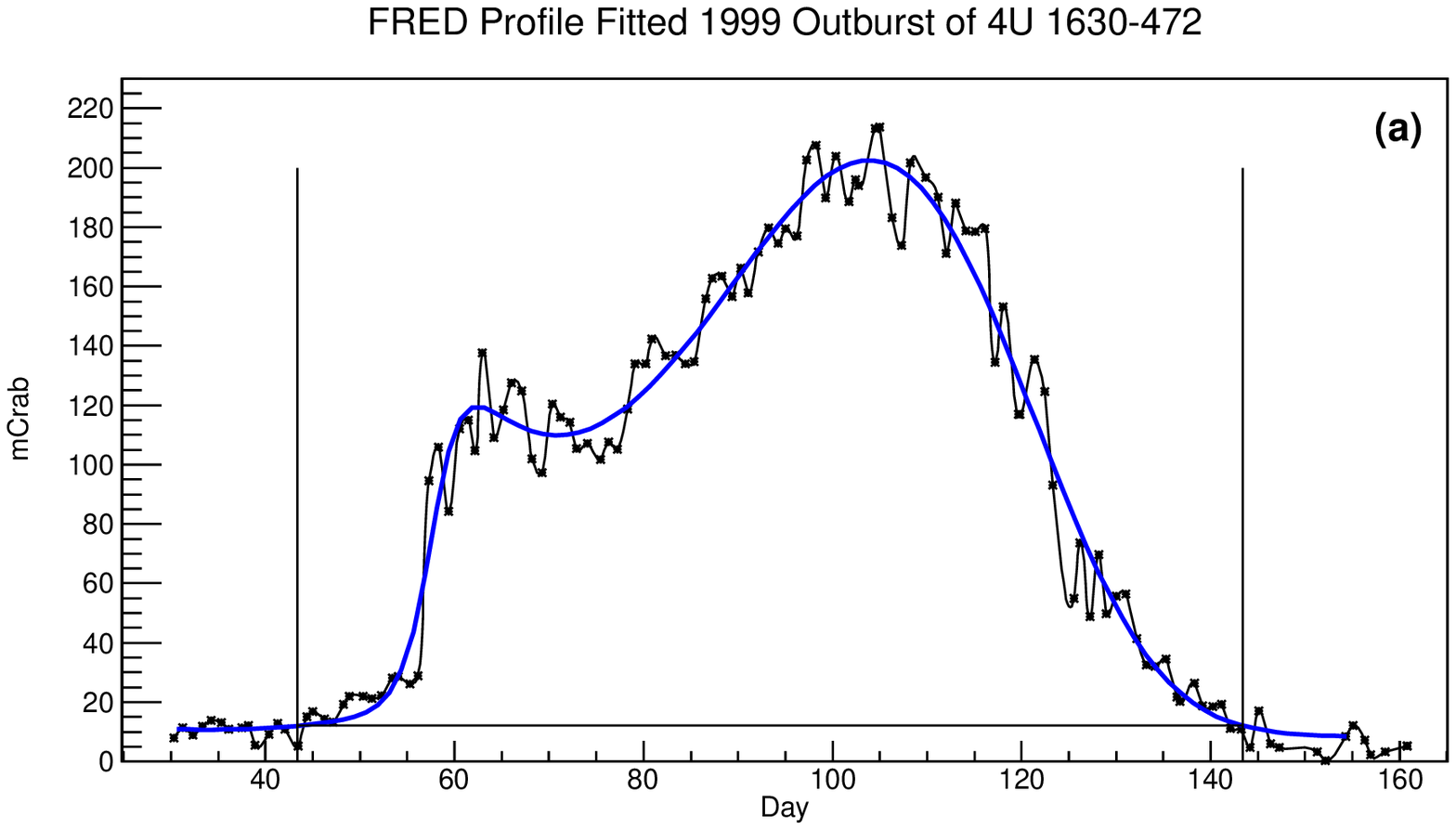}\hskip 0.5cm
\includegraphics[width=8.0truecm, height=4cm]{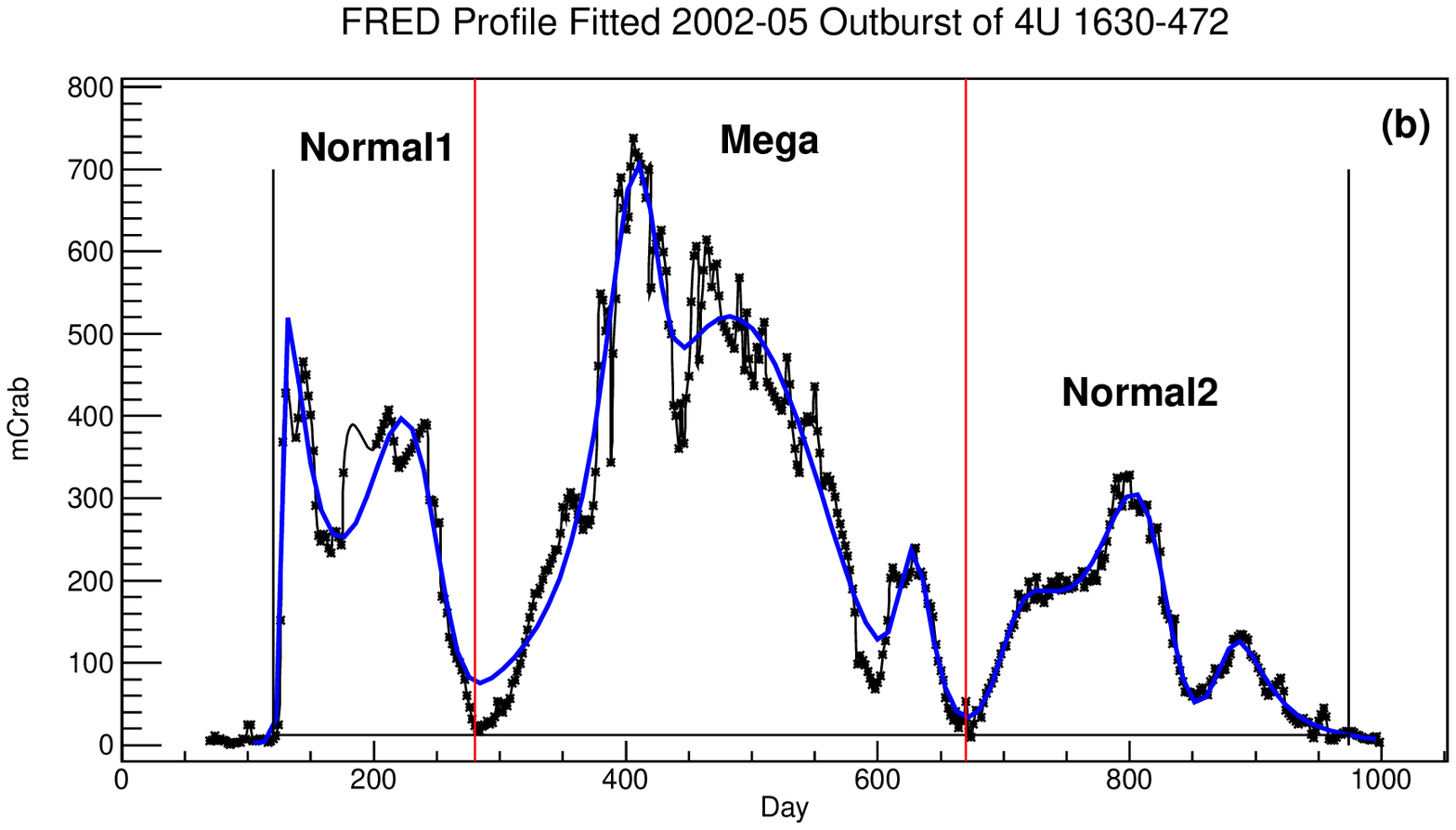}}
\vskip 0.1cm
\hskip 0.5cm
\vbox{
\includegraphics[width=8.0truecm, height=4cm]{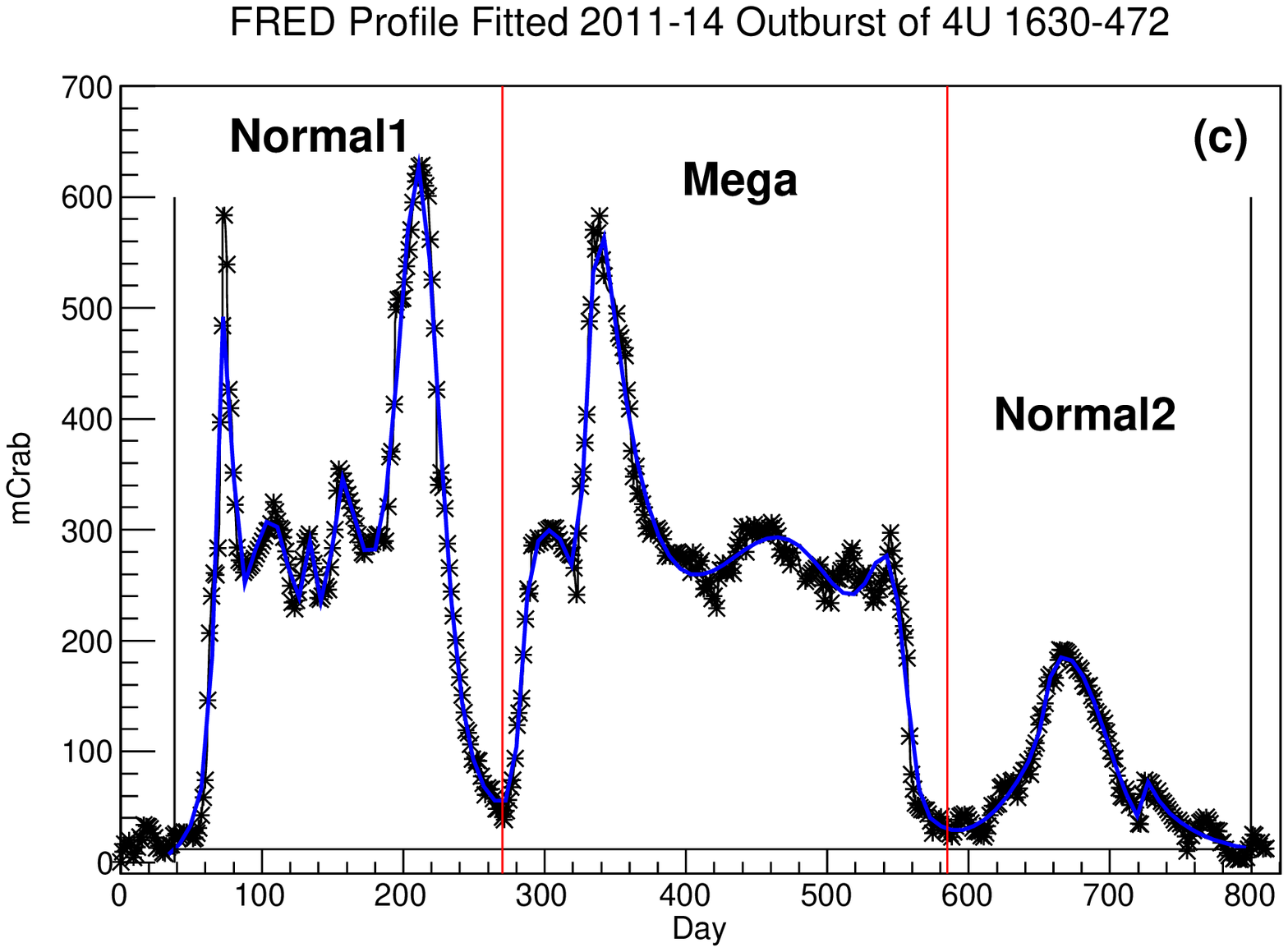}\hskip 0.5cm
\includegraphics[width=8.0truecm, height=4cm]{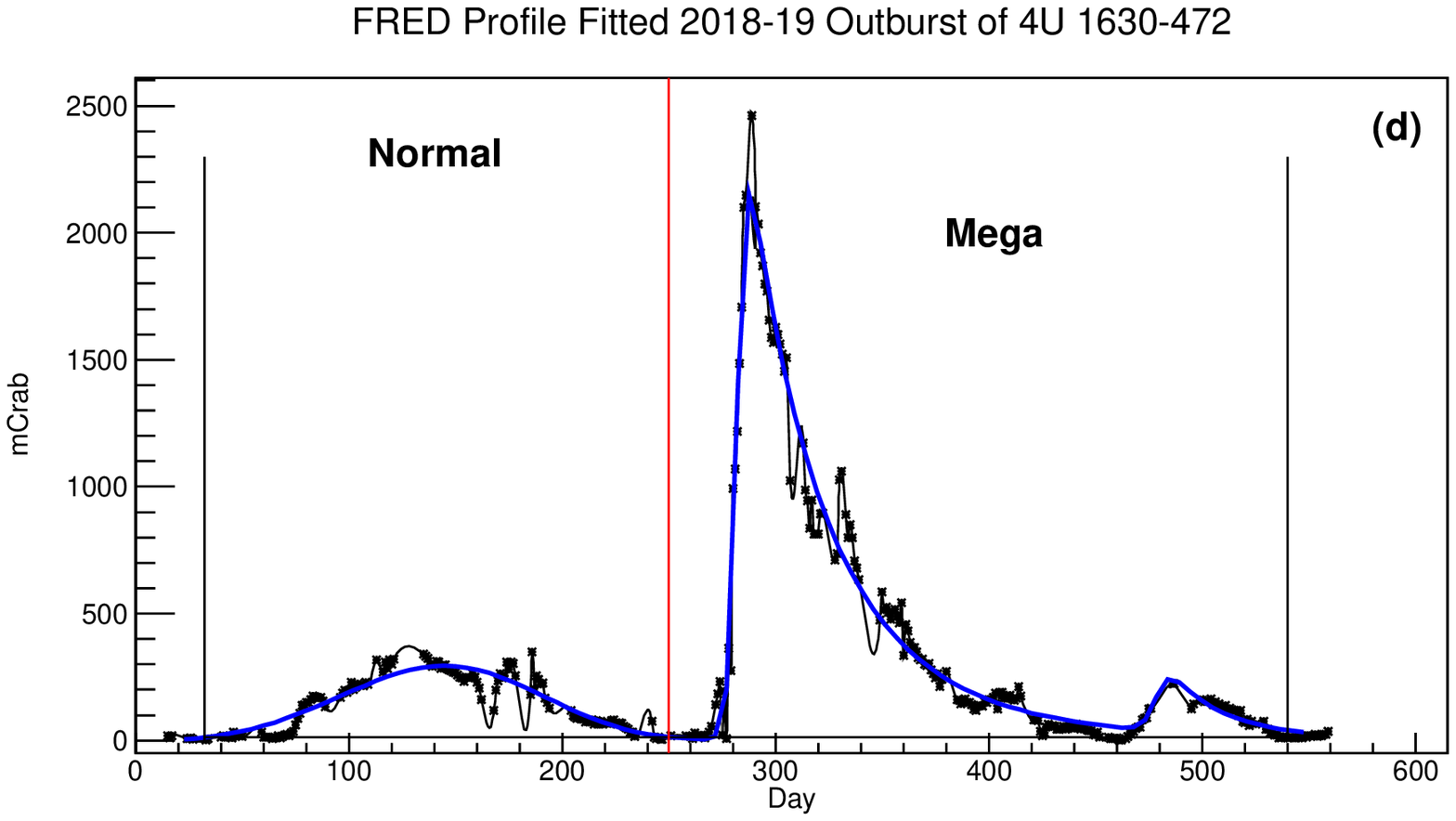} }
\vskip 0.1cm
    \caption{FRED profile fitted (a) 1999 normal and (b) 2002--05, (c) 2011--14, (d) 2018--19 super outbursts. A $12~mCrab$ line (black) is also drawn to signify
                    the start and end of the outburst in the rising and declining phases respectively. Solid red lines are used to differentiate between normal and super
                    outbursts.}
\end{figure}

\subsection{Calculation of Integrated Fluxes}

An outburst profile contain one or multiple peaks having nature of fast-rise and exponential decay. So, we use the FRED model to fit each peak of the outburst. 
The combined FRED fit allows us to get overall nature of the outburst profiles. The equation of the FRED model (Kocevski et al. 2003) is defined as,

$$F(t) = F_m\left(\frac{t}{t_m}\right)^r\left[\frac{d}{d+r}+\frac{r}{d+r}\left(\frac{t}{t_m}\right)^{r+1}\right]^{-\frac{r+d}{r+1}} \eqno (1)$$

where, $F_m$ is called the peak flux, i.e., the highest flux value in the light curve, $t_m$ is the peak time, i.e., the time at which the source or
the outburst reaches its maximum flux value. $r$ and $d$ signify the rising and decaying indices. One thing we have to keep in mind is that, here 
$r$ and $d$ do not give the value of the slopes in the rising and declining phases.  

We have also used numerical simple integration method (trapezoidal to be more specific) to estimate integral flux during the outburst. This method allows us 
to verify the integral flux calculated using the FRED model fitting. We have integrated the fluxes of each outbursts from their respective daily average light 
curves using this method and normalized them using the $2009-10$ outburst flux from MAXI/GSC data. We notice that all the integrated fluxes using this method 
lie very close to the values integrated by FRED model method (see, Table 1). However, the downloaded daily average light curves have some discrepancies. 
Some of the data have negative values as well as for some day, the data are missing. There is also some random variations in the data where it shows sudden 
dips or ups. These are mainly due to the instrumental error. Since there is no significant difference between the two methods, it is more scientific to make 
use of the FRED model. In Bhowmick et al. (2021), we have made use of both these methods for our analysis of the BHC GX 339-4. 

As stated above, in the first method, we have fitted all the light curves using FRED model, as shown for the $2009-10$ outburst in Figure 2. Using the 
Bayesian Information Criterion (BIC) selection method, we confirmed that $k$-number of FRED profiles that are needed to obtain best fit of the entire 
lightcurve profile of an outburst (for more details see, Bhowmick et al. 2021). Same selection method is followed for all the outbursts of 4U 1630-472. 
We have estimated the integral fluxes (I.F.) for all these outbursts from the model fits using the combined FRED profiles. Assuming $12~mCrab$ flux as 
outburst threshold (Chakrabarti et al. 2019), we first mark the start and end of an outburst using first and last occurrence of the $12~mCrab$ flux points 
in the rising and declining phases respectively. We show the variation of the estimated integral fluxes in Figure 3. 
Here in panel (a), we show the normalized flux per day during the outbursts and in panel (b), the normalized flux per day in case of accumulation
periods. The accumulation period of each outburst is the time during which matter accumulate at a distance, far away from the black hole 
(namely pile-up radius $X_p$) before the outburst (Chakrabarti et al. 2019). The width of each bar in panel (a) signifies the duration of that particular 
outburst. We have used the MAXI/GSC $2009-10$ outburst fitted I.F. value to normalize all the I.F. values as it is less noisy than the RXTE/ASM one. 

As mentioned in the previous Section, 4U~1630-472 showed two different nature (normal and super) of outbursts. The super outbursts are long duration and 
their light curve profiles show that they contain one or two normal outbursts. Normal outbursts are roughly periodic in their outburst duration and 
accumulation periods (quiescent period prior to an outburst). From these two parameters and light curve profiles, we found that 2002--05 and 2011--14 
super outburst contained two normal and one mega outbursts, where as 2018--19 super outburst contained one mega and one normal outbursts (see, Fig. 2).
The 2002--05 super outburst started on MJD $52521.4 \pm 0.4$ after a quiescent gap of close to 460 days after the previous 2000--01 outburst. 
If we look closely into this 2002--05 outburst, we can see that, after passing through rise and decay phases within $\sim 170$~days, the flux became 
as low as $\sim 12$~mCrab close to MJD $\sim$ 52680, which is close to the average duration of a normal outburst. If this would have been a normal 
outburst, then the next normal outburst would have occurred close to the time where we saw third peak of the 2002--05 super outburst. 
This third peak started close to MJD $\sim$ 53080. Near to this day, the flux became very low ($\sim 12$ mCrab) before again starting to rise. 
This third peak also had a duration, close to the average duration of normal outbursts. The 2nd peak was observed between MJD $\sim$ 52680 and 
MJD $\sim$ 53080, which continued for a longer duration than its other two peaks. We termed this as the `mega' outburst. We recalculated the integral 
fluxes for these three outbursts separately from the combined FRED model fit and using start and end time of each outbursts as days mentioned above. 
The 1st and 3rd peaks are termed as 2002--03 and 2004 normal outburst respectively and 2nd peak is termed as 2003--04 mega outburst. 
A similar procedure is followed to separate normal outbursts from mega outbursts within the super outbursts for 2011--14 and 2018--19. Integral 
fluxes are also calculated by selecting suitable inputs of start and end dates of the normal and mega outbursts, where flux became close 
to $\sim 12$ mCrab within the super outbursts.

\begin{figure*}
  \centering
    \includegraphics[angle=0,width=10cm,keepaspectratio=true]{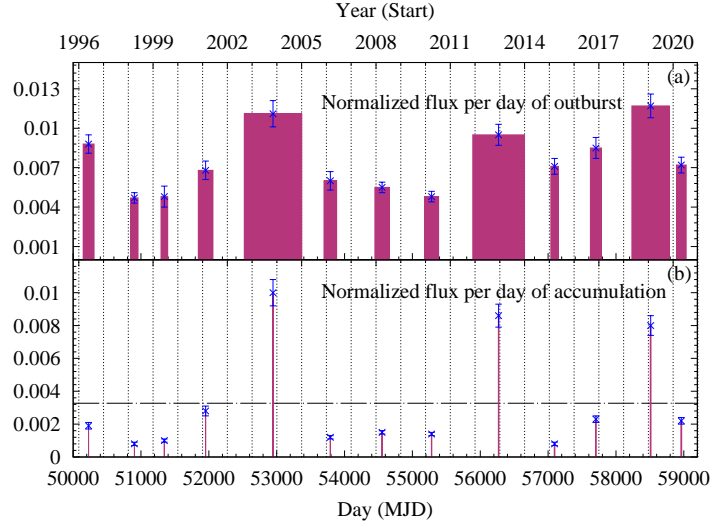}
    \caption{Histogram plot of normalized flux per day for (a) outburst and (b) accumulation periods. All these estimations are done using the $2009-10$
             values.}
\end{figure*}

\begin{figure*}
  \centering
  \begin{minipage}[]{0.45\textwidth}
    \includegraphics[width=\textwidth]{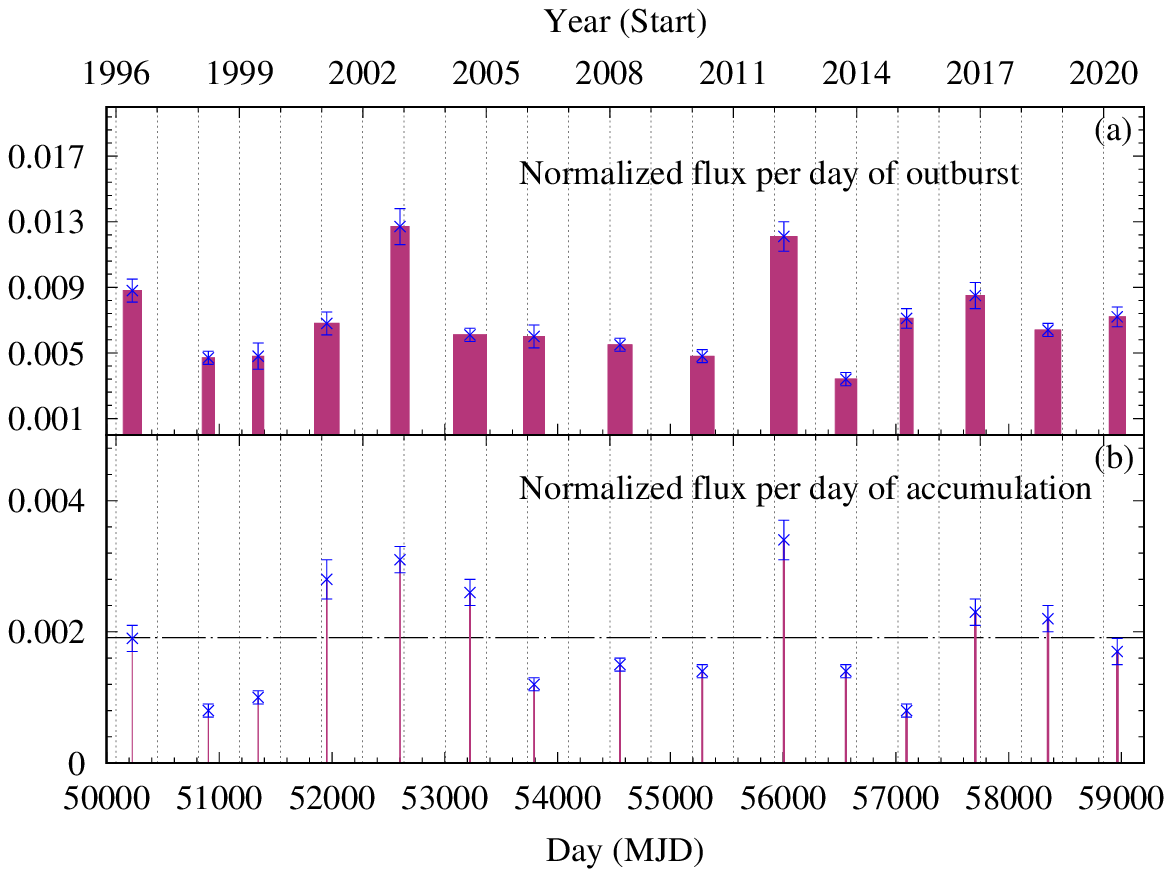}
    \caption{Histogram plot of normalized flux per day for (a) outburst and (b) accumulation periods in case of only normal outbursts. All these estimations 
             are done using the $2009-10$ values.}
\end{minipage}
  \hfill
  \begin{minipage}[]{0.45\textwidth}
    \includegraphics[width=\textwidth]{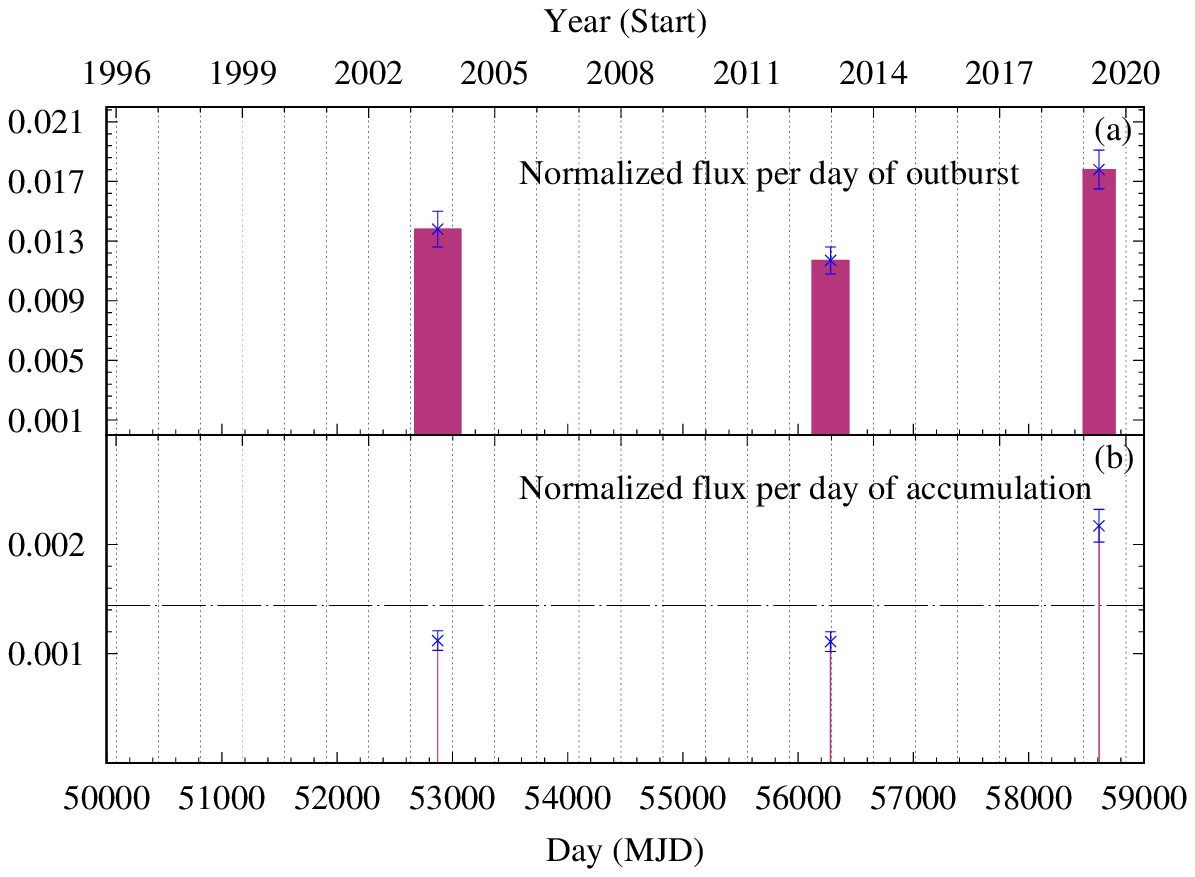}
    \caption{Histogram plot of normalized flux per day for (a) outburst and (b) accumulation periods in case of only mega outbursts. All these estimations 
             are done using the $2009-10$ values.}
\end{minipage}
\end{figure*}

\begin{figure*}
\vskip 1.2cm
  \centering
  \begin{minipage}[b]{0.45\textwidth}
    \includegraphics[width=\textwidth]{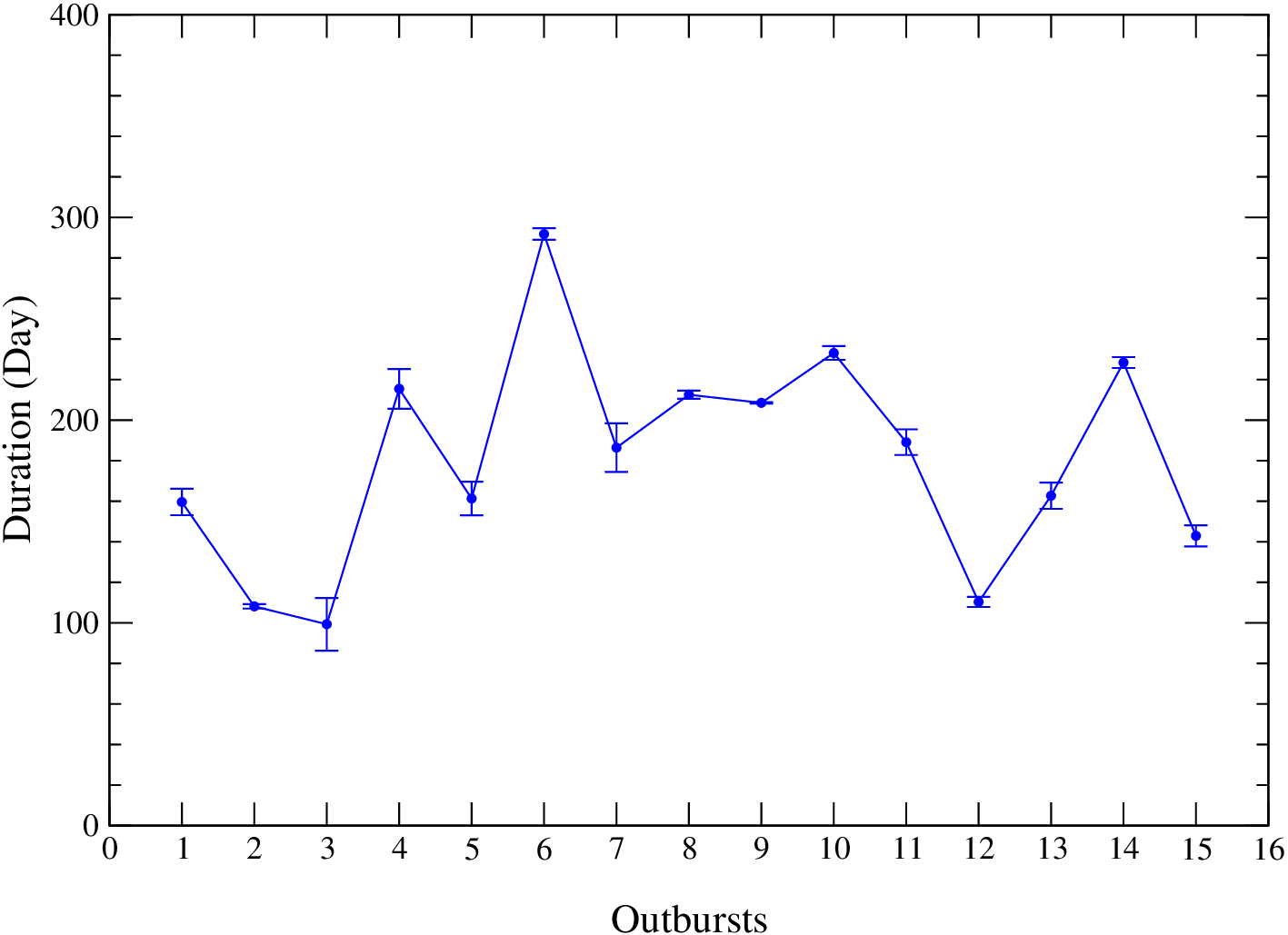}
    \caption{Variation of duration between normal outbursts. Here, each tick represents the number of normal outbursts.}

\end{minipage}
  \hfill
  \begin{minipage}[b]{0.47\textwidth}
    \includegraphics[width=\textwidth]{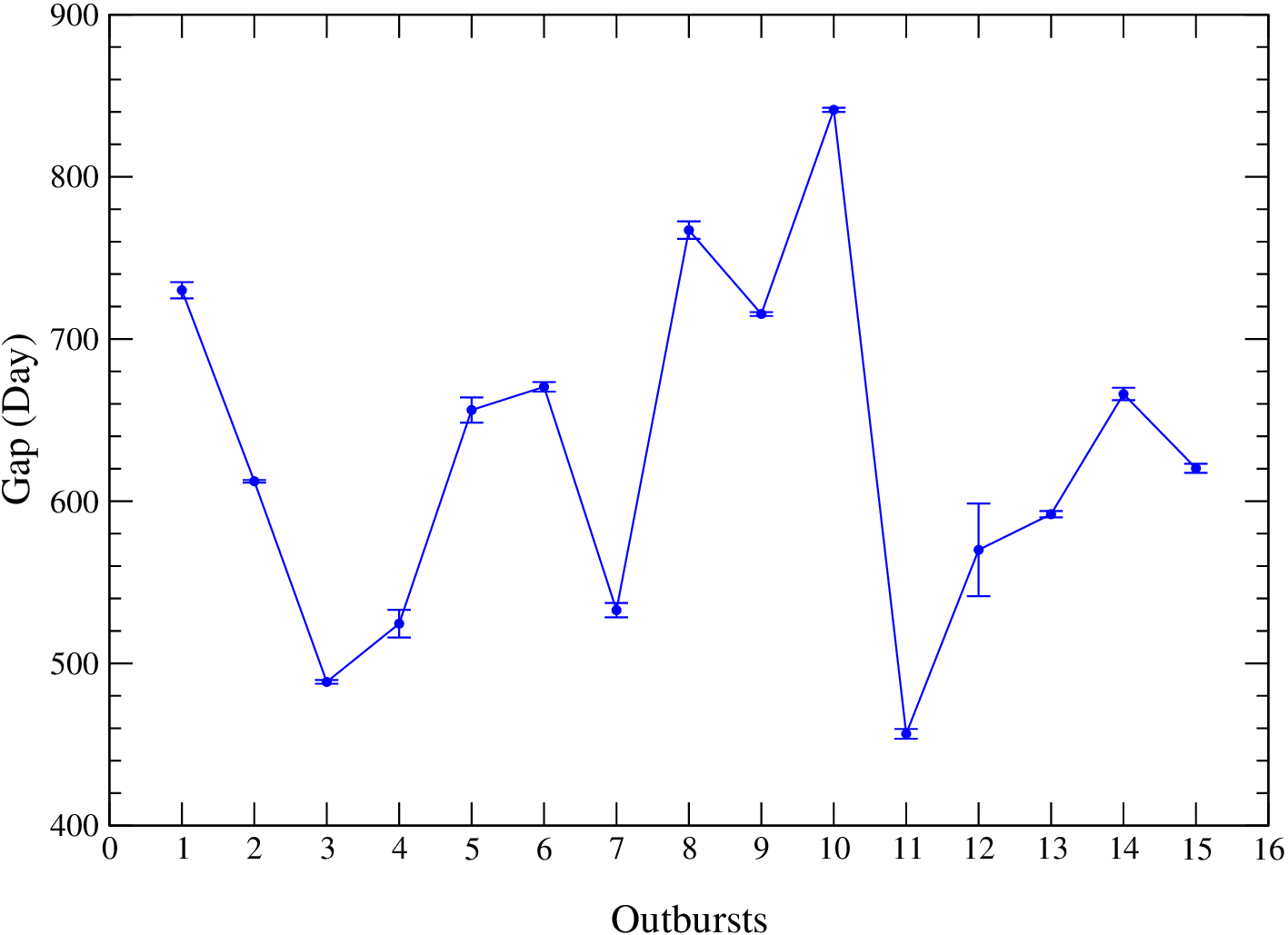}
    \caption{Variation of gaps between normal outbursts. Here, each tick represents the number of normal outbursts.}
\end{minipage}
\end{figure*}

\begin{figure*}
\vskip 0.8cm
  \centering
  \begin{minipage}[b]{0.45\textwidth}
    \includegraphics[width=\textwidth]{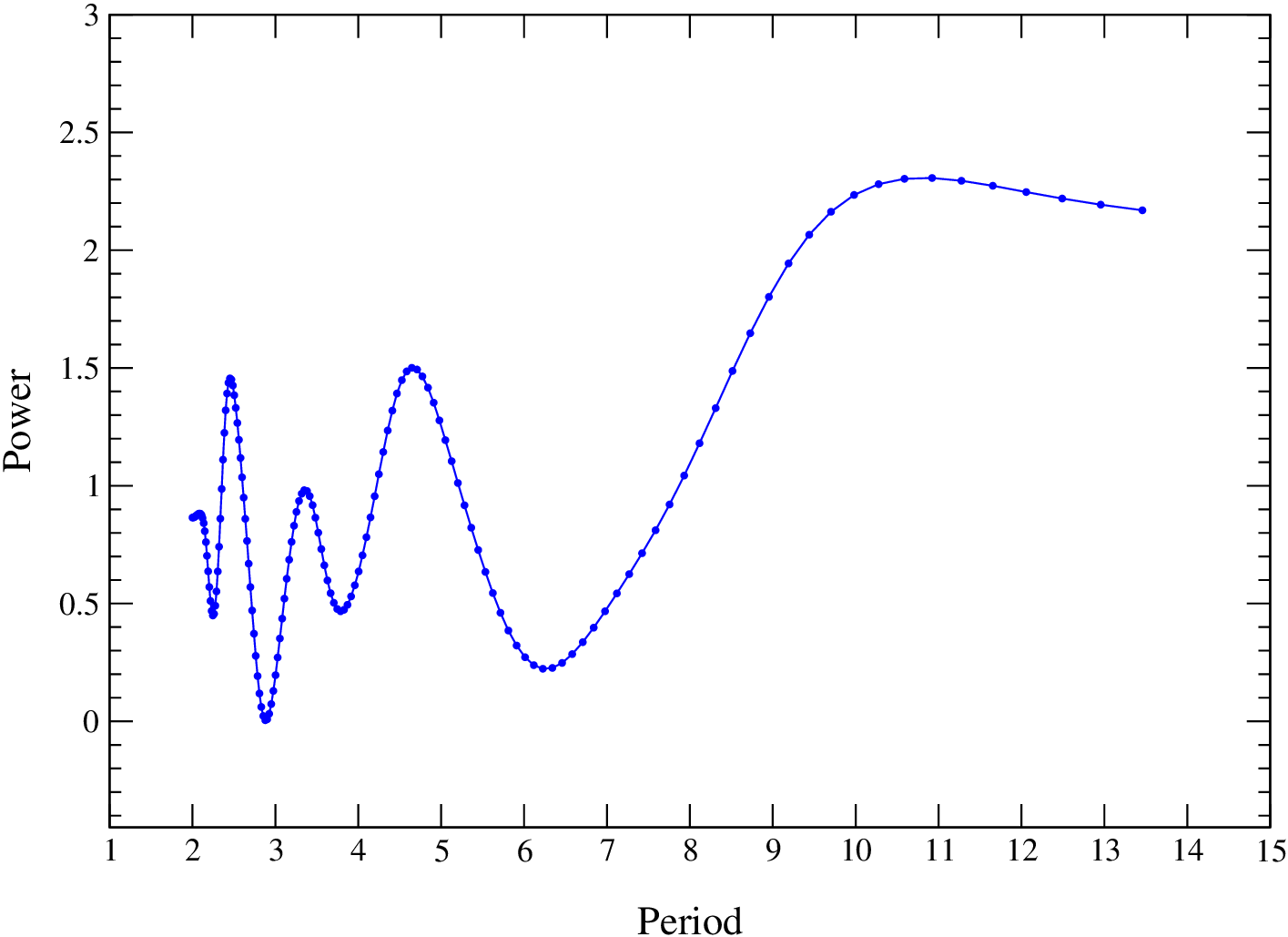}
    \caption{Lomb Scargle Periodogram plot for periodicity calculation of duration between normal outbursts (Figure 6).}
             
\end{minipage}
  \hfill
  \begin{minipage}[b]{0.45\textwidth}
    \includegraphics[width=\textwidth]{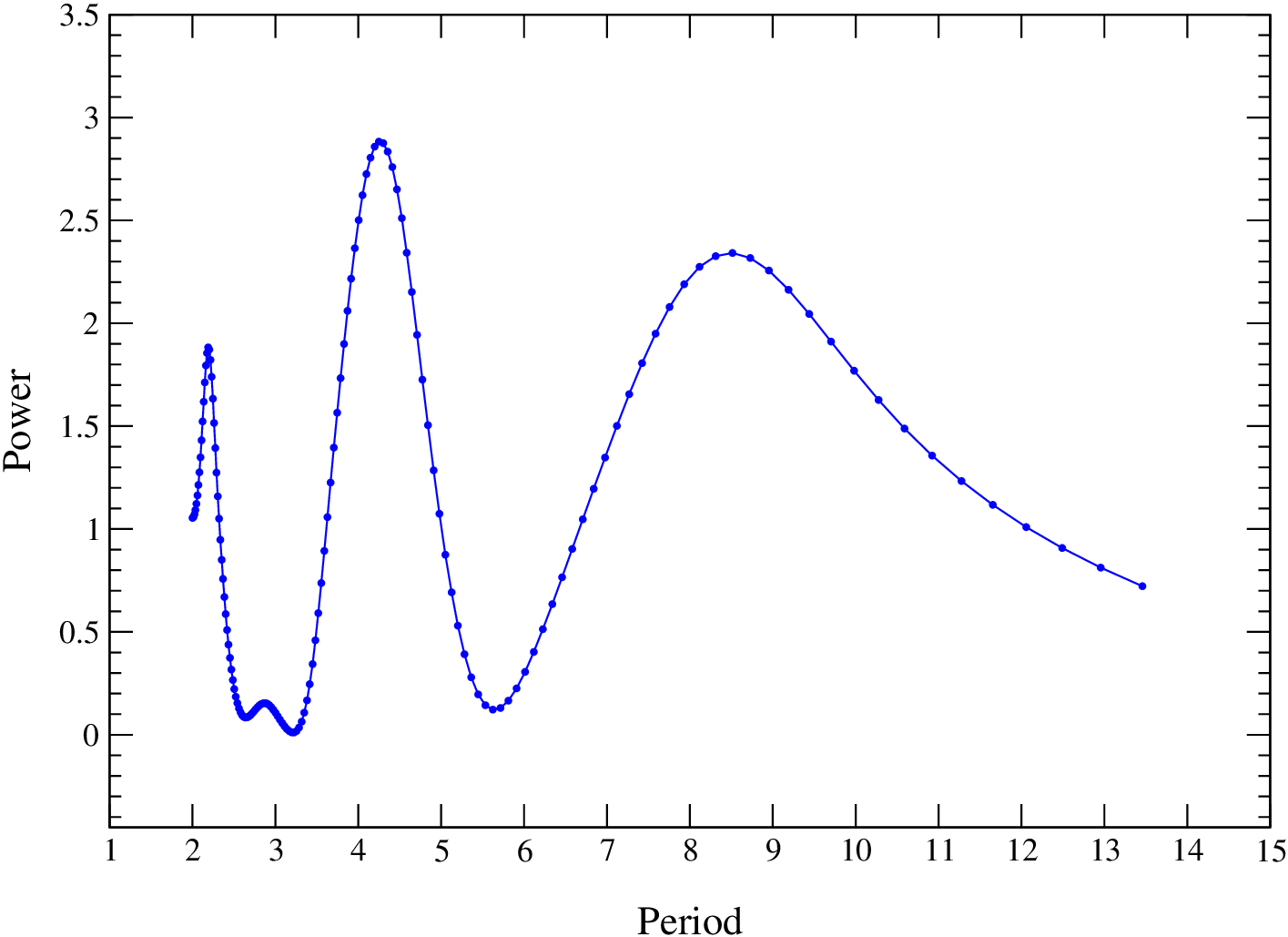}
    \caption{Lomb Scargle Periodogram plot for periodicity calculation of gaps between normal outbursts (Figure 7).}
\end{minipage}
\end{figure*}

\subsection{Comparative Study of Outburst and their Fluxes}

To have a comparative study of all the outbursts of 4U 1630-472, observed by two different satellite (RXTE and MAXI) instruments (ASM and GSC), we make 
use of commonly observed outburst i.e., $2009-10$ outburst fitted parameters to normalize integral flux and other parameters of the other outbursts. 
In Fig. 3, a histogram is drawn, where we show variation of the normalized integral fluxes per day of outburst or accumulation periods of all 
outbursts including both the normal and super outbursts. The evolution of the same integral fluxes for the 15 normal and 3 mega outbursts, after their 
separation from the super outbursts whenever applicable are shown in Figs. 4 and 5 respectively. Average integral fluxes for per day of accumulation i.e. 
quiescent periods prior to each outburst are marked with the horizontal dashed lines in each Figures. 
 
Figure 3(a) shows the integrated flux per day during the time of the outburst. This is shown for all the outbursts using a histogram, where the width 
of each histogram signifies the duration of every outburst. So, the area of each histogram represents the total I.F. during that outburst. These integral 
fluxes represent the total energy released during the outbursts (normalized with respect to the $2009-10$ outburst I.F.). We notice that three outbursts 
($2002-05$, $2011-14$ and $2018-19$) have higher values of I.F. during outbursts. Figure 3(b) shows the integrated flux per day in the period of matter 
accumulation prior to each outburst. The black dashed line, that is drawn in this panel of the Figure, represents the average of all these values. 
This actually tells us about the mass transfer from the pile up radius $X_p$. When this value is high, i.e., for the cases of $2002-05$, $2011-2014$ and 
$2018-19$, it means that more matter is coming from the $X_p$. This also tells us that the more matter was accumulated prior to the triggering of these 
three outbursts. That is why, the values for these outbursts are very high and much above the average values. These are termed as the super outbursts, 
in which the I.F. for both the cases of outburst and accumulation is higher than the other short period outbursts. As we see in the other outbursts, I.F. 
is much less and stays roughly close to each other. This tells us that during these normal outbursts, the supply of matter from the companion is not as 
large as the cases of those super outbursts. Except for the presence of the super outbursts, the nature of 4U~1630-472 seems fairly the same as of other 
SXTs. In those sources also, the supply of matter from the companion also changes from time to time for different outbursts. The presence of these super 
outbursts makes the SXT 4U 1630-472 an interesting candidate to study. 

In the previous Section, we have discussed the method about separation of the super outbursts into its two constituents : normal and mega. In Figure 4(a), 
we show the normalized integrated flux per day during outburst for only the normal outbursts including the normal parts of the super outbursts, while 
Figure 4(b) shows that for accumulation period. If we look in panel (b), we can notice that during the accumulation period the normalized flux per day 
for all the outbursts stay close to an average value. Figure 5(a-b) shows the same properties for only the mega parts of the super outbursts. We see that 
while the accumulation of matter were close to each other for 2003--04 and 2012--13 outbursts, it increased for the 2018--19 one. All the estimated 
properties are given in Table 1, 2 and 3 for the cases of all the outbursts, only normal outbursts and only mega outbursts respectively.

\vskip 1.0cm
\subsection{Estimation of Periodicity of Duration and Gap of Normal Outbursts}

We have first estimated the duration and gaps between normal outbursts from our FRED fitted light curve profiles. Duration of outbursts are the 
time periods between start and end of 12 mCrab fluxes. Gaps between outbursts are the time from the peak of the preceding outburst to the peak of the 
current outburst. Here, normal outbursts also includes the normal parts from the super outbursts. We have plotted the variations of these two properties 
in Figures 6 and 7 respectively. These show quasi periodic type of variations. This is why we tried to estimate their periodicity.

We have mainly used the Lomb Scargle Periodogram (LSP) method from \href{https://exoplanetarchive.ipac.caltech.edu/cgi-bin/Pgram/nph-pgram}
{NASA exoplanet archive} to estimate these periodicities. The LSP is a variation of the Discrete Fourier Transform (DFT), which is optimized to identify 
sinusoidal shaped signals in time series data. If there is a sinusoidal signal of with a frequency $\omega_0$, the periodogram gives a large peak value 
of power at the period $T=2\pi/\omega_0$. When the data is not purely sinusoidal (like in our case), the x-axis value of the largest peak in power 
signifies the period of the data. The NASA exoplanet archive implemented the periodogram power as the normalization of the total power ($P_x(\omega)$) 
by variance ($\sigma$) or $P_n(\omega)=P_x(\omega)/\sigma^2$, where $P_n(\omega)$ is the normalized power. We show the LSP plots for normal outburst 
duration and gap in Figures 8 and 9 respectively, i.e., Figures 8 and 9 corresponds to Figures 6 and 7 respectively. For our case, we have obtained 
highest power values of $\sim$ 2.31 and 2.88 for normal outburst duration and gaps respectively. These values correspond to periods of 10.92 and 4.25 
respectively. This tells us that the duration and gap of normal outbursts has periodicities of 10.92 and 4.25 respectively.

\subsection{Contamination of Closely Situated Pulsars}

\begin{figure}[h!]
  \centering
    \includegraphics[angle=0,width=08cm,keepaspectratio=true]{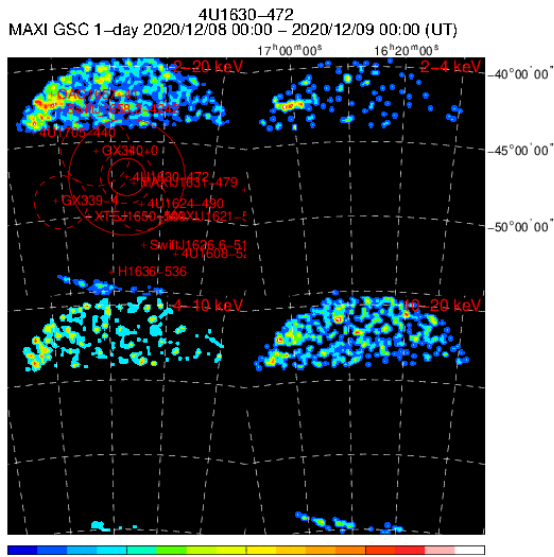}
    \caption{One day (from 2020 December 08 00:00 to 2020 December 09 00:00) MAXI/GSC FOV image of 4U 1630-472 in different energy bands. This image 
             is downloaded from the \href{http://maxi.riken.jp}{MAXI} website.} 
\end{figure}

To check for the contamination from the nearly situated IGR pulsar (IGR J16358-4726, IGR J16393-4643, IGR J16320-4751) sources, we have also analyzed their 
light curve profiles. For IGR J16358-4726, we use Swift/BAT light curve in 15-50 keV energy band (since ASM and GSC data are not available), for IGR J16393-4643 
we use archival 15-50 keV Swift/BAT and 1.5-12 keV RXTE/ASM light curves (since GSC data is not available), while for IGR J16320-4751, we only use the 1.5-12 
keV RXTE/ASM (as BAT and GSC data are not available). Using all the proper conversion factors, we have converted the light curves into mCrab unit and fitted 
them with FRED model. During the outbursts of 4U 1630-472, none of these sources were very much active and average flux was always close to our assumed 
threshold value ($\sim$ 12 mCrab) for quiescent.

\section{Discussion and Concluding Remarks}
 
The soft X-ray transient (SXT) 4U 1630-472 was discovered by \textit{Uhuru} satellite. This SXT is a very active source and showed recurring outbursts 
in a gap of 400--550 days since its discovery. Nature of the outbursts of the source are studied here in RXTE and post RXTE era. Daily average archival 
data of RXTE/ASM ($1.5-12$ keV) and MAXI/GSC ($2-10$ keV) are used for our study. Outburst profiles are fitted with the FRED model, to extract peak flux, 
integral flux, duration, etc. parameters. Here, we would like to discuss about some anomalous nature of the source and then try to draw concluding remarks. 

One of the goals of this paper is to study the triggering mechanism of an outburst. Although there is debate on it, we believe that it could be due 
to the sudden rise in viscosity (Ebisawa et al. 1996) due to increase in rate of supply of matter toward black hole via Roche lobe overflow or wind. Recently, 
Chakrabarti, Debnath and Nagarkoti (2019; CDN19) and Bhowmick et al. (2021) studied the nature of outbursts of the recurring transient BHCs H~1743-322 and 
GX~339-4 during the RXTE and MAXI (present) eras. It is suggested by CDN19 that, there is a pile-up radius ($X_p$) which is situated at a distance, far away 
from the source where matter, which comes in from the companion, at first starts to accumulate due to low viscosity. Then when a large amount of matter sets 
in at $X_p$, the temperature of the disk rises which increases the instability of the disk. As a result, viscosity of the accumulated matter starts to rise. 
When viscosity attains a critical value, matter rushes towards the BH, triggering an outburst. When this distance is large, more matter accumulates at $X_p$ 
and more time is needed to rise the viscosity to critical value to start an outburst. When this $X_p$ is closer, it takes less time to start the outburst. 
The outburst duration, which depends on the amount of matter accumulated from the companion, is indirectly dependent on the location of $X_p$ also. 
This rise in viscosity could also be due to the rise in magnetic activity of the companion. When matter brings in magnetic field from the companion, 
it can remove angular momentum by ejecting piled up matter. As a result, matter rushes towards the compact source.

We can notice that, unlike other transient sources, the outbursts of 4U 1630-472 are different to some extent. First of all, the outbursts are not of 
similar nature. There are outbursts which lasted for $100-200$ days, defined as the normal outbursts. There are also few outbursts which lasted for 
1.5--2.5 years. We term them as the super outbursts. This type of nature of outbursts is generally not common in case of other BHCs. From our detailed 
analysis of the super outbursts, we found that the super outbursts are the combination of two types of outbursts: mega and normal outbursts. Here, we 
make an effort to separate contribution of the normal and mega outbursts from each of the super outbursts. The FRED model fitted parameters of each 
outbursts (normal, super and mega) are noted in Tables 1-3. We use 2009--10 outburst, commonly observed by both the RXTE and MAXI satellites, to normalize 
integral fluxes of all the outbursts. In Fig. 3, histogram plots of estimated fluxes per day of the outburst and per day of the accumulation period of all 
the 13 observed outbursts of this source are shown. Similar histogram plots of the normal and mega outbursts are plotted in Figs. 4 \& 5.

From the detailed analysis of the duration of the outburst and quiescent phases, we observe that the recurrence period (both accumulation and outburst) 
of the outbursts has changed over the years. Although normal outbursts are observed $\sim 500$ days of gap, their integral fluxes of the outbursts have
varied roughly in a cyclic way (Fig. 4). From Figures 6 and 7, we can also notice the cyclic variations of outburst and accumulation duration
of normal outbursts. The Lomb Scargle Periodogram of these variations have helped us find the periodicities for these two duration. For outburst and
accumulation duration (i.e., gap), we have found period of 10.92 and 4.25 respectively. However, in case of mega outbursts, we observed decreasing nature 
in both the outburst duration and the accumulation period between the outbursts. From the decreasing trend of accumulation period of mega outbursts and 
average recurrence period of normal outbursts, we can predict that the next super outburst will consists of one normal and one mega outbursts.

Generally, during an outburst of a BHC, the changes of spectral states form a hysteresis loop or in case of a failed outburst it stays mostly in 
harder states (Remillard \& McClintock 2006). For 4U 1630-472, most of the outbursts show anomalies in case of hard state (HS) to high soft state (HSS) 
transition (Capitanio et al. 2015). There is lack of hard state in both the rising and declining phases (checked from hardness ratio or HR). In some 
outbursts, the final detection level of HS is very close to the quiescent luminosity (Capitanio et al. 2015). We have seen for H 1743-322 (Chakrabarti et 
al. 2019; Fig. 5c) and GX 339-4 (Bhowmick et al. 2021; Fig. 3c) that there is roughly constant matter supply from the companion. However, for this source, 
the matter supply seems to be varying in a quasi periodic way (see Fig. 4(b)) for different normal outbursts. All these together make the 4U 1630-472 
source an interesting and peculiar candidate in terms of outbursting nature. We think that quasi periodic nature of matter supply for normal outbursts is 
linked with the location of the pile up radius ($X_p$). The supply of matter could be constant for all the outbursts. However, $X_p$ is changing its 
location in a cyclic way. When it is closer to the BH, less amount of matter needs to be accumulated to start an outburst and accumulation is higher when 
$X_p$ is away. For example, before the 1996 outburst matter was accumulating at a farther $X_p$ before triggering an outburst. This is why the average 
amount of accumulated matter is high. For the next outburst in 1997--98, $X_p$ was on a closer location and hence the average accumulated matter is less. 
We think that this change of location of $X_p$ have taken place in a quasi periodic way which reflects our results in Figure 4. For the cases of mega 
outbursts also, we think that the pile-up location became closer gradually. i.e., for the 2003--04 mega one $X_p$ was farther and the outburst gap as well 
as the duration was higher. For 2012--13 case, the $X_p$ was at a closer location and thus the accumulation gap was less as well as the duration. In 
2018--19, it came more close as both the duration and gap both decreased. For this outburst a huge amount of matter rushed towards the BH and this outburst 
reached a very high peak, which is not similar with the other outbursts. This cyclic change in $X_p$ location could be one of the plausible explanations 
for the nature of normal outbursts of this source if the supply of matter from the companion remains same over the years. For mega outbursts, closing of
the pile-up location is one factor which describes the duration and also the gap between them. There could be another factor why these mega outbursts are
taking place.

Searching for contamination from nearby IGR pulsars, we checked their activity during the active outburst phases of 4U 1630-472. We have not found 
any significant contribution from these pulsars to the outbursts of this source. Till now, there is no complete information on the companion of this source. 
If we look at the MAXI/GSC field of view (FOV) of the source, we can find that there are many sources in close region where 4U 1630-472 is situated (Figure 
10). The second reason behind the occurrence of mega outbursts could be that the source is in a three body (or trinary) system. The black hole is in a binary 
configuration and there could be another source, coupled to this binary system. Due to this configuration, there could be an extra unknown perturbation which 
is working on the system against which the binary could stay stable. However, due to this extra perturbation, there could be extra supply of matter from the 
companion when the third body comes in a closer distance. Due to this, there is an extra amount of radiation, coming in as the mega outbursts.

\section{Summary}

We have studied the outburst profiles of the BHC 4U 1630-472 using archival daily average light curve data from RXTE/ASM (1.5--12 keV) and MAXI/GSC 
(2--10 keV) satellites. There are two types of outbursts in this SXT: normal outbursts, which lasts for 100--200 days and super outbursts, lasting for 
1.5--2.5 years. We have assumed that the super ones are the combination of normal ones and an extra part, which we term as `mega' outbursts. The average 
recurrence period for normal outbursts also supports this assumption of the occurrence of normal outbursts inside super outbursts. We have separated the 
contributions of these normal and mega outbursts from the super outbursts. We have fitted those light curve profiles (converted to flux into mCrab unit) 
using the fast rise exponential decay (FRED) model. Using the resultant FRED fitted profile and also numerical (Trapezoidal rule) method, we have estimated 
integral fluxes from all the outbursts. From the FRED fitted profile, we have extracted some of the other outburst properties (peak flux, peak day, duration, 
normalized flux per day for the cases of both outburst and accumulation period etc) for this BHC. Based on these analysis, we have tried to explain the 
nature of its outbursts. 

If the matter transfer from the companion is not changing much then we think that the pile-up radius ($X_p$) is changing in a cyclic manner. 
Before some outbursts it is away from the source, while before some outbursts it is closer to the source. This is why the average flux in the 
accumulation period looks quasi periodic in nature. For mega outbursts also $X_p$ was closer than the previous one and thus resulting in decrement 
in the outbursts duration as well as the accumulation gap.

There could be a third body, coupled to the binary system which is causing an extra perturbation. As a result, the mass transfer is occasionally 
changing on a very large scale which is causing the mega outbursts.

\section*{Acknowledgement}

This work has made use of ASM data of NASA's RXTE satellite and GSC data provided by RIKEN, JAXA and the MAXI team.
K.C. acknowledges support from DST/INSPIRE (IF170233) fellowship. 
D.D. acknowledges support from Govt. of West Bengal, India and ISRO sponsored RESPOND project (ISRO/RES/2/418/17-18) fund.
R.B. acknowledges support from CSIR-UGC NET qualified UGC fellowship (June-2018, 527223).
S.K.N. and D.D. acknowledge partial support from ISRO sponsored RESPOND project (ISRO/RES/2/418/17-18) fund.

\newcommand{\STAB}[1]{\begin{tabular}{@{}c@{}}#1\end{tabular}}

\begin{table*}
 \addtolength{\tabcolsep}{-3.5pt}
 \centering
 \caption{FRED model fitted parameters in case of all the outbursts}
 \label{tab:table1}
\begin{tabular}{cccccccccccccc}
\hline
& Outburst    &        Duration    &     Peak Flux        &      Peak Day        &     Accumulation   & Normalized Integrated  & Normalized Integrated  \\
&   Year      &                    &                      &                      &     Period         & Flux Per Day during    & Flux Per Day during    \\
&	     &         (Days)     &      (mCrab)         &      (MJD)           &     (Days)$^a$     &   Outburst $^b$        &   Accumulation $^b$    \\
\hline
       & RXTE/ASM&&&&&&\\
& $  1996   $ &  $ 159.6 \pm 6.5 $ & $ 296.711 \pm 11.6 $ & $ 50253.1 \pm 0.8 $  & $ 730.1 \pm 5.0 $  &  $ 0.0088 \pm 0.0007 $ & $ 0.0019 \pm 0.0002 $  \\
& $ 1997-98 $ &  $ 108.1 \pm 1.1 $ & $ 382.088 \pm 16.3 $ & $ 50865.4 \pm 0.1 $  & $ 612.3 \pm 0.8 $  &  $ 0.0047 \pm 0.0004 $ & $ 0.0008 \pm 0.0001 $  \\
& $  1999   $ &  $ 99.30 \pm 13  $ & $ 203.100 \pm 12.8 $ & $ 51354.0 \pm 1.1 $  & $ 488.6 \pm 1.2 $  &  $ 0.0048 \pm 0.0008 $ & $ 0.0010 \pm 0.0001 $  \\
& $ 2000-01 $ &  $ 215.4 \pm 9.8 $ & $ 596.026 \pm 35.8 $ & $ 51878.5 \pm 7.4 $  & $ 524.5 \pm 8.6 $  &  $ 0.0068 \pm 0.0007 $ & $ 0.0028 \pm 0.0003 $  \\
& $ 2002-05 $ &  $ 847.2 \pm 22  $ & $ 704.987 \pm 13.4 $ & $ 52811.6 \pm 0.3 $  & $ 933.1 \pm 7.7 $  &  $ 0.0111 \pm 0.0010 $ & $ 0.0100 \pm 0.0008 $  \\
& $  2005   $ &  $ 186.4 \pm 12  $ & $ 277.561 \pm 13.3 $ & $ 53738.0 \pm 4.8 $  & $ 926.4 \pm 4.4 $  &  $ 0.0060 \pm 0.0007 $ & $ 0.0012 \pm 0.0001 $  \\
& $ 2007-08 $ &  $ 212.5 \pm 2.0 $ & $ 217.831 \pm 7.80 $ & $ 54505.1 \pm 0.6 $  & $ 767.1 \pm 5.4 $  &  $ 0.0055 \pm 0.0004 $ & $ 0.0015 \pm 0.0001 $  \\
\hline
        & \multicolumn{1}{c}{MAXI/GSC}&&&&&&\\
& $ 2009-10 $ &  $ 208.5 \pm 0.3 $ & $ 316.117 \pm 15.4 $ & $ 55220.5 \pm 0.6 $  & $ 715.4 \pm 1.2  $ &  $ 0.0048 \pm 0.0004 $ & $ 0.0014 \pm 0.0001 $  \\
& $ 2011-14 $ &  $ 761.5 \pm 22  $ & $ 627.838 \pm 13.5 $ & $ 56061.8 \pm 5.9 $  & $ 841.3 \pm 30.7 $ &  $ 0.0095 \pm 0.0008 $ & $ 0.0086 \pm 0.0007 $  \\
& $ 2014-15 $ &  $ 110.3 \pm 2.5 $ & $ 346.898 \pm 9.80 $ & $ 57088.4 \pm 1.8 $  & $ 1026.6\pm 28.6 $ &  $ 0.0071 \pm 0.0006 $ & $ 0.0008 \pm 0.0001 $  \\
& $ 2016-17 $ &  $ 162.7 \pm 6.5 $ & $ 316.027 \pm 7.80 $ & $ 57680.4 \pm 1.6 $  & $ 592.0 \pm 1.9  $ &  $ 0.0085 \pm 0.0008 $ & $ 0.0023 \pm 0.0002 $  \\
& $ 2018-19 $ &  $ 555.3 \pm 8.2 $ & $ 2152.46 \pm 35.3 $ & $ 58489.1 \pm 0.2 $  & $ 808.7 \pm 1.7  $ &  $ 0.0117 \pm 0.0009 $ & $ 0.0080 \pm 0.0006 $  \\
& $  2020   $ &  $ 142.9 \pm 5.2 $ & $ 396.872 \pm 21.7 $ & $ 58966.8 \pm 2.2 $  & $ 477.7 \pm 2.8  $ &  $ 0.0072 \pm 0.0006 $ & $ 0.0022 \pm 0.0002 $  \\ 
\hline
\end{tabular}
\vskip 0.2cm
 \noindent{
\leftline{$^a$ Accumulation period is the duration between peak days from the previous to the current outburst.}
\leftline{$^b$ This normalization is done by the integrated flux value of the $2009-10$ outburst.}
}
\end{table*}

\begin{table*}
\small
 \addtolength{\tabcolsep}{-3.5pt}
 \centering
 \caption{FRED model fitted parameters for only normal outbursts}
 \label{tab:table2}
\begin{tabular}{cccccccccccccc}
\hline
 Outburst    &        Duration    &     Peak Flux        &      Peak Day        &     Accumulation   & Normalized Integrated  & Normalized Integrated  \\
   Year      &                    &                      &                      &     Period         & Flux Per Day during    & Flux Per Day during    \\
	     &         (Days)     &      (mCrab)         &      (MJD)           &     (Days)$^a$     &   Outburst $^b$        &   Accumulation $^b$    \\
\hline
        RXTE/ASM&&&&&&\\
 $  1996   $ & $ 159.6 \pm 6.5  $ & $ 296.711 \pm 11.6 $ & $ 50253.1 \pm 0.8  $ & $ 730.1 \pm 5.0  $ & $  0.0088 \pm 0.0007 $ & $ 0.0019 \pm 0.0002 $  \\
 $ 1997-98 $ & $ 108.1 \pm 1.1  $ & $ 382.088 \pm 16.3 $ & $ 50865.4 \pm 0.1  $ & $ 612.3 \pm 0.8  $ & $  0.0047 \pm 0.0004 $ & $ 0.0008 \pm 0.0001 $  \\
 $  1999   $ & $ 99.30 \pm 13   $ & $ 203.100 \pm 12.8 $ & $ 51354.0 \pm 1.1  $ & $ 488.6 \pm 1.2  $ & $  0.0048 \pm 0.0008 $ & $ 0.0010 \pm 0.0001 $  \\
 $ 2000-01 $ & $ 215.4 \pm 9.8  $ & $ 596.026 \pm 35.8 $ & $ 51878.5 \pm 7.4  $ & $ 524.5 \pm 8.6  $ & $  0.0068 \pm 0.0007 $ & $ 0.0028 \pm 0.0003 $  \\
 $ 2002-03 $ & $ 161.3 \pm 8.3  $ & $ 536.422 \pm 11.9 $ & $ 52534.7 \pm 2.3  $ & $ 656.2 \pm 7.8  $ & $  0.0127 \pm 0.0011 $ & $ 0.0031 \pm 0.0002 $  \\
 $  2004   $ & $ 291.8 \pm 2.8  $ & $ 308.743 \pm 8.41 $ & $ 53205.2 \pm 1.9  $ & $ 670.5 \pm 2.9  $ & $  0.0061 \pm 0.0004 $ & $ 0.0026 \pm 0.0002 $  \\
 $  2005   $ & $ 186.4 \pm 12   $ & $ 277.561 \pm 13.3 $ & $ 53738.0 \pm 4.8  $ & $ 532.8 \pm 4.4  $ & $  0.0060 \pm 0.0007 $ & $ 0.0012 \pm 0.0001 $  \\
 $ 2007-08 $ & $ 212.5 \pm 2.0  $ & $ 217.831 \pm 7.80 $ & $ 54505.1 \pm 0.6  $ & $ 767.1 \pm 5.4  $ & $  0.0055 \pm 0.0004 $ & $ 0.0015 \pm 0.0001 $  \\

\hline
        MAXI/GSC&&&&&&\\
 $ 2009-10 $ & $ 208.5 \pm 0.3  $ & $ 316.117 \pm 15.4 $ & $ 55220.5 \pm 0.6  $ & $ 715.4 \pm 1.2  $ & $  0.0048 \pm 0.0004 $ & $ 0.0014 \pm 0.0001 $  \\ 
 $ 2011-12 $ & $ 233.1 \pm 3.4  $ & $ 627.838 \pm 7.27 $ & $ 56061.8 \pm 1.1  $ & $ 841.3 \pm 1.3  $ & $  0.0121 \pm 0.0009 $ & $ 0.0034 \pm 0.0003 $  \\ 
 $ 2013-14 $ & $ 189.1 \pm 6.3  $ & $ 185.407 \pm 2.31 $ & $ 56518.4 \pm 2.8  $ & $ 456.6 \pm 3.0  $ & $  0.0034 \pm 0.0004 $ & $ 0.0014 \pm 0.0001 $  \\ 
 $ 2014-15 $ & $ 110.3 \pm 2.5  $ & $ 346.898 \pm 9.80 $ & $ 57088.4 \pm 1.8  $ & $ 570   \pm 28.6 $ & $  0.0071 \pm 0.0006 $ & $ 0.0008 \pm 0.0001 $  \\ 
 $ 2016-17 $ & $ 162.7 \pm 6.5  $ & $ 316.027 \pm 7.80 $ & $ 57680.4 \pm 1.6  $ & $ 592.0 \pm 1.9  $ & $  0.0085 \pm 0.0008 $ & $ 0.0023 \pm 0.0002 $  \\ 
 $  2018   $ & $ 228.4 \pm 2.7  $ & $ 293.998 \pm 4.70 $ & $ 58346.5 \pm 3.4  $ & $ 666.1 \pm 3.8  $ & $  0.0064 \pm 0.0004 $ & $ 0.0022 \pm 0.0002 $  \\
 $  2020   $ & $ 142.9 \pm 5.2  $ & $ 396.872 \pm 21.7 $ & $ 58966.8 \pm 2.2 $  & $ 620.3 \pm 3.2  $ & $  0.0072 \pm 0.0006 $ & $ 0.0017 \pm 0.0002 $  \\ 

\hline
\end{tabular}
\vskip 0.2cm
 \noindent{
\leftline{$^a$ and $^b$ represent the same things as Table 1.}
}
\end{table*}

\begin{table*}
\small
 \addtolength{\tabcolsep}{-3.5pt}
 \centering
 \caption{FRED model fitted parameters for only mega outbursts}
 \label{tab:table3}
\begin{tabular}{cccccccccccccc}
\hline
 Outburst    &        Duration    &     Peak Flux        &      Peak Day        &     Accumulation   & Normalized Integrated  & Normalized Integrated  \\
   Year      &                    &                      &                      &     Period         & Flux Per Day during    & Flux Per Day during    \\
	     &         (Days)     &      (mCrab)         &      (MJD)           &     (Days)$^a$     &   Outburst $^b$        &   Accumulation $^b$    \\
\hline
        RXTE/ASM&&&&&&\\
 $ 2003-04 $ & $ 403.0 \pm 1.2 $  & $ 704.987 \pm 13.4 $ & $ 52811.6 \pm 0.3 $  & $ 4943.9 \pm7.72 $ & $ 0.0138 \pm  0.0012 $ & $ 0.00112 \pm 0.00009 $ \\

\hline
        MAXI/GSC&&&&&&\\
 $ 2012-13  $ & $ 320.0 \pm 7.7  $ & $ 627.838 \pm 13.5 $ & $ 56061.8 \pm 5.9  $ & $ 3378.7 \pm30.7  $ & $ 0.0117 \pm 0.0009 $ & $ 0.00111 \pm0.00009 $ \\
 $ 2018-19  $ & $ 280.0 \pm 5.2  $ & $ 2152.46 \pm 35.3 $ & $ 58489.1 \pm 0.2  $ & $ 2298.8 \pm1.71  $ & $ 0.0178 \pm 0.0013 $ & $ 0.00217 \pm0.00015 $ \\
\hline
\end{tabular}
\vskip 0.2cm
 \noindent{
\leftline{$^a$ and $^b$ represent the same things as Table 1.}
}
\end{table*}


\begin{thebibliography}{99}

\bibitem[]{} Bhowmick, R., Debnath, D., Chatterjee, K., et al., 2021, ApJ, 910, 138 
\bibitem[]{} Capitanio, F., Campana, R., De Cesare, G., Ferrigno, C., 2015, MNRAS, 450, 3840
\bibitem[]{} Casella, P., Belloni, T., Stella L., 2005, ApJ, 629, 403
\bibitem[]{} Chakrabarti, S. K., Debnath, D., Nagarkoti, S., 2019, AdSpR, 63, 3749
\bibitem[]{} Chen, W., Shrader, C. R., Livio, M., 1997, ApJ, 491, 312
\bibitem[]{} Debnath, D., Chakrabarti, S. K., \& Nandi, A., 2010, A\&A, 520, 98
\bibitem[]{} Debnath, D., Chakrabarti, S. K., Nandi, A., 2013, AdSpR, 52, 2143
\bibitem[]{} Debnath, D., Jana, A., Chakrabarti, S. K., Chatterjee, D., Mondal, S., 2017, ApJ, 850, 52
\bibitem[]{} Ebisawa, K., Titarchuk, L., Chakrabarti, S. K., 1996, PASJ, 48, 59
\bibitem[]{} Horne, J. H., \& Baliunas, S. L., 1986, ApJ, 302, 757
\bibitem[]{} Jones, C., Forman, W., Tananbaum, H., Turner, M.J.L., 1976, ApJ 210, L9 
\bibitem[]{} Kocevski, D., Ryde, F., Liang, E., 2003, ApJ, 596, 389
\bibitem[]{} Kovács, G., Zucker, S., \& Mazeh, T., 2002, A\&A, 391, 369
\bibitem[]{} Kuulkers, E., Parmar, A. N., Kitamoto, S., Cominsky, R. L., Sood, R. K., 1997, MNRAS, 291, 81
\bibitem[]{} Nandi, A., Debnath, D., Mandal, S., Chakrabarti, S. K., 2012, A\&A, 542, A56
\bibitem[]{} Parmar, A. N., Williams, O. R., E. Kuulkers, E., Angelini, L., White, N. E., 1997, A\&A, 319, 855
\bibitem[]{} Plavchan, P., Jura, M., Kirkpatrick, J. D., Cutri, R. M., \& Gallagher, S. C., 2008, ApJSS, 175, 191
\bibitem[]{} Priedhorsky, W.C., 1986, AsSS, 126, 89
\bibitem[]{} Remillard, R. A., McClintock, J. E., 2006, ARA\&A, 44, 49
\bibitem[]{} Scargle, J. D., 1982, ApJ, 263, 835
\bibitem[]{} Seifina, E., Titarchuk, L., Shaposhnikov, N., 2014, ApJ, 789, 57
\bibitem[]{} van der Klis, M., 2005, AN, 326, 798


\end{thebibliography}
\end{document}